\documentclass{jfm}

\usepackage{amsmath}    
\usepackage{graphicx}   
\usepackage{verbatim}   
\usepackage{color}      
\usepackage{subfigure}  
\usepackage{hyperref}   
\usepackage{tensor}
\usepackage{amsfonts}
\usepackage{natbib}
\usepackage{amssymb}

\newcommand{\bnabla}{\boldsymbol{\nabla}}
\newcommand{\gsim}{\lower.7ex\hbox{$\;\stackrel{\textstyle>}{\sim}\;$}}
\newcommand{\lsim}{\lower.7ex\hbox{$\;\stackrel{\textstyle<}{\sim}\;$}}
\newcommand{\la}{\left<}
\newcommand{\ra}{\right>}

\title[Two-dimensionalization of low-$Rm$ turbulence ]{Exact two-dimensionalization of low-magnetic-Reynolds-number flows subject to a strong magnetic field}
\author[B. Gallet and C. R. Doering]{Basile Gallet$^1$ \and Charles R. Doering$^2$}
\affiliation{$^1$Service de Physique de l'\'Etat Condens\'e, DSM, CNRS UMR 3680, CEA Saclay, 91191 Gif-sur-Yvette, France\\[\affilskip]
$^2$Department of Physics, Department of Mathematics, and Center for the Study of Complex Systems, University of Michigan, Ann Arbor, MI 48109, USA}

\begin{document}

\maketitle

\begin{abstract}
We investigate the behavior of flows, including turbulent flows,  driven by a horizontal body-force and subject to a vertical magnetic field, with the following question in mind: for very strong applied magnetic field, is the flow mostly two-dimensional, with remaining weak three-dimensional fluctuations, or does it become {\it exactly} 2D, with no dependence along the vertical? 

We first focus on the quasi-static approximation, i.e. the asymptotic limit of vanishing magnetic Reynolds number $Rm \ll 1$: we prove that the flow becomes exactly 2D asymptotically in time, regardless of the initial condition and provided the interaction parameter $N$ is larger than a threshold value. We call this property {\it absolute two-dimensionalization}: the attractor of the system is necessarily a (possibly turbulent) 2D flow.

We then consider the full-magnetohydrodynamic equations and we prove that, for low enough $Rm$ and large enough $N$, the flow becomes exactly two-dimensional in the long-time limit provided the initial vertically-dependent perturbations are infinitesimal. We call this phenomenon {\it linear two-dimensionalization}: the (possibly turbulent) 2D flow is an attractor of the dynamics, but it is not necessarily the only attractor of the system. Some 3D attractors may also exist and be attained for strong enough initial 3D perturbations.

These results shed some light on the existence of a dissipation anomaly for magnetohydrodynamic flows subject to a strong external magnetic field.

\end{abstract}

\section{Introduction}


Liquid metal turbulence is encountered in various situations, ranging from metallurgy \citep{Davidson} to the flow in the Earth's outer core \citep{Moffatt}, including laboratory experiments on magnetohydrodynamic (MHD) turbulence \citep{Alemany,Gallet}. Because their kinematic viscosity is much lower than their magnetic diffusivity, with magnetic Prandtl numbers $Pm$ typically in the range of $10^{-6}$ to $10^{-5}$, flows of liquid metals can be turbulent and still be characterized by a small magnetic Reynolds number: in most laboratory experiments on MHD turbulence, the kinetic Reynolds number is in the range $10^4-10^6$, with a magnetic Reynolds number rarely exceeding unity. At small enough scales, turbulence in Earth's outer core may also be considered as a low-$Rm$ flow.

Turbulence subject to a strong external magnetic field tends to become two-dimensional, the velocity field becoming weakly dependent on the coordinate along the applied magnetic field (hereafter denoted as the vertical axis by convention). Indeed, any flow that depends on the vertical coordinate shears the applied field and induces electrical currents. For low magnetic Reynolds number, such currents are rapidly damped through Ohmic dissipation and the flow structures that depend on the vertical are more damped than vertically-invariant eddies. This intuitive argument can be made more precise in the asymptotic limit of vanishing magnetic Reynolds number $Rm$, the so-called quasi-static approximation. In this limit the applied field simply appears as an additional Ohmic damping term in the  Navier-Stokes equation \citep{Sommeria, Knaepen}. This term is dominant for large $B_0$, and it depends on the second vertical derivative of the velocity: for infinite $B_0$ the bulk velocity field must be independent of the vertical.

For finite but strong external magnetic field $B_0$, one may ask the following questions:  does the bulk flow remain three-dimensional, with weak three-dimensional velocity fluctuations superposed on a strong two-dimensional flow, or does it become purely two-dimensional? How do the Ohmic and viscous dissipation rates compare for large $B_0$? Is there a scaling law relating the vertical and horizontal gradients of the velocity field? Or is there a critical value of $B_0$ above which the flow becomes completely independent of the vertical?

These central questions are strongly related to the scaling of the energy dissipation in MHD turbulence. Indeed, three dimensional hydrodynamic turbulence displays a dissipation anomaly: the 3D flow of a fluid with kinematic viscosity $\nu$ forced on a scale $\ell$ with typical velocity $U$ dissipates a finite power per unit mass $\epsilon \sim U^3/\ell$ as the Reynolds number $U \ell / \nu$ goes to infinity. By contrast, body-forced two-dimensional flows do not present such a singular zero-viscosity limit: in 2D the dissipated power vanishes for vanishing viscosity, $\ell$ and $U$ being kept constant. Because the MHD situation with strong external $B_0$ is an intermediate situation between 2D and 3D turbulence, an open question is whether it displays a dissipation anomaly.

In the laboratory, the influence of an external magnetic field on the bulk turbulence of a liquid metal was studied in various geometries, ranging from channel flows to confined turbulence \citep{Alemany, Gallet, Klein, Potherat2014}. These studies indicated a clear trend toward two-dimensionalization as the external field increases. Experiments  were also performed in shallow layers of liquid metal subject a strong uniform magnetic field to maximize the two-dimensionalization \citep{Sommeria86, GalletGAFD}. The corresponding flows displayed good agreement with the phenomenology of 2D turbulence.

Numerical simulations at unit magnetic Prandtl number have also displayed partial two-dimensionalization, with the formation of large-scale two-dimensional structures coexisting with small-scale three-dimensional fluctuations \citep{Alexakis2D,Bigot}. \citet{Zikanov} focused their numerical study on MHD turbulence in the quasi-static limit $Rm \to 0$, and they observed three regimes: for weak applied field, the flow is three-dimensional. For intermediate $B_0$, its behavior is intermittent, with long phases of quasi-2D motion interrupted by 3D bursts. For large $B_0$, the system settles into a purely 2D state, the amplitude of the $z$-dependent part of the velocity field decreasing exponentially in time. Although the forcing procedure is somewhat unrealistic (they rescale the total energy of the large-scale Fourier modes to a constant value at each time step), their simulations provide a clean example of MHD turbulence reaching a purely 2D stationary state provided the external field is strong enough. On the theoretical side, they consider a given two-dimensional large-scale cellular flow, and they show that it is absolutely stable with respect to vertically-dependent perturbations for large enough external field. In a subsequent paper \citep{Thess}, they study the linear stability of several other steady two-dimensional flows as a function of the strength of the external field. For finite Reynolds number, both the elliptical vortex and the vortex sheet are stable provided the applied field is strong enough. Although enlightening, these examples do not give an unambiguous answer to the central question asked above, because they do not consider the general setup of body-forced turbulence.

Insightful analytical results on the behavior of such body-forced MHD flows have been obtained by \citet{Potherat2003} in the vanishing-$Rm$ approximation: using rigorous methods from mathematical analysis, these authors provide upper bounds on the dimension of the attractor of MHD turbulent flows. For strong external magnetic field, the bound follows the same scaling as the one obtained (with the same method) for purely 2D flows. This indicates that the behavior of such low-$Rm$ MHD flows somewhat ``resembles" that of 2D flows, but it does not prove that the flow really is 2D: for instance, the bound holds even when the forcing is three-dimensional, a situation for which the flow cannot be purely two-dimensional.

In this paper we use rigorous analysis and estimates to answer the central question of exact two-dimensionalization. We consider an MHD flow driven by a horizontal body force and subject to a vertical magnetic field, in a domain that is periodic in the horizontal and bounded vertically by stress-free surfaces. Such boundary conditions are chosen to remove the boundary layers inherent to experimental realizations, and to allow for a purely two-dimensional velocity field. We stress the fact that the results presented in this paper carry over to the 3D periodic cube, as discussed in the final section. In the quasi-static ($Rm\to 0$) approximation, we prove that the flow becomes exactly 2D asymptotically in time, regardless of the initial condition, provided the interaction parameter $N$ is larger than a critical value $N_c(Re)$. We call this property {\it absolute two-dimensionalization}. We then consider the full-MHD system and study the evolution of vertically-dependent perturbations of infinitesimal amplitude. We show that such infinitesimal perturbations decrease provided the interaction parameter is larger than some $Re$-dependent threshold value, and the magnetic Reynolds number is lower some other $Re$-dependent threshold. We thus extend the domain in which we prove that the two-dimensional flow is an attractor, but at the price of considering infinitesimal 3D perturbations only. We therefore call this property {\it linear two-dimensionalization}.

In section \ref{section2} we describe the MHD system and recall some bounds on the viscous energy dissipation in hydrodynamic turbulence. Section \ref{section3} contains the proof of absolute two-dimensionalization for quasi-static MHD ($Rm \to 0$). Sections \ref{section4} and \ref{section5} deal with linear two-dimensionalization for the full MHD equations. In section \ref{section6} we summarize the different results obtained in this paper and discuss their implications for the energy dissipation rate of low-$Rm$ MHD turbulence. Some  technical aspects of the proofs are presented in three appendices.

The rigorous estimates of this study are based on Young's, H\"older's, Poincar\'e's, the Cauchy-Schwarz, and Ladyzhenskaya's inequalities. Young's inequality is simply the fact that $|ab|\leq \frac{1}{2c}a^2+\frac{c}{2}b^2$ for any $c>0$, where $c$ can be optimized to get the best possible upper bound. H\"older's, Poincar\'e's and the Cauchy-Schwarz inequalities are presented in a fluid dynamics context in \citet{Doering}, while Ladyzhenskaya's inequality is proven in \citet{Ladyzhenskaya} (Lemma 1, page 8).


\section{Body-forced magnetohydrodynamics and hydrodynamics\label{section2}}

\subsection{Magnetohydrodynamics in a cubic domain}

\begin{figure}
\begin{center}
\includegraphics[width=120 mm]{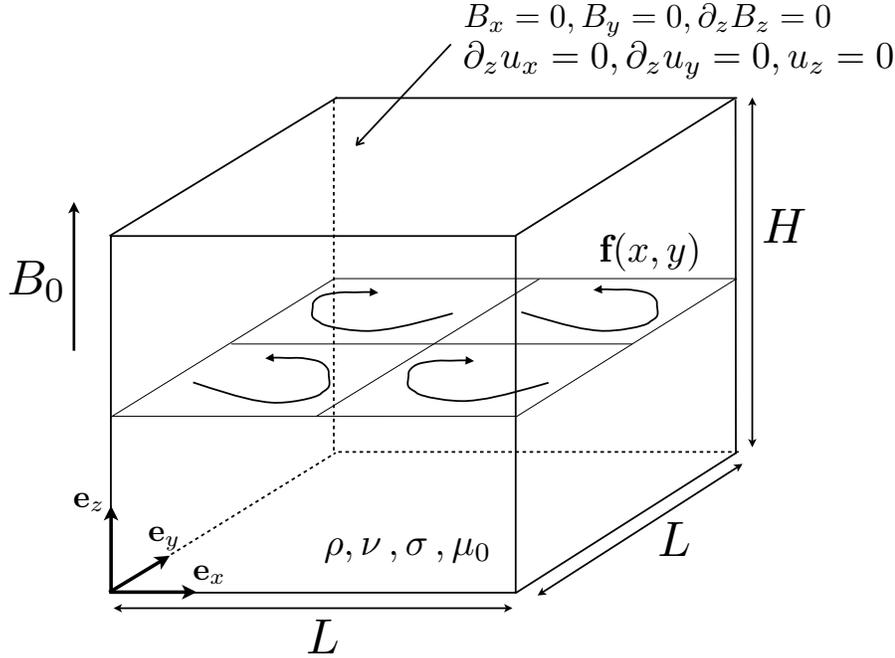}
\end{center}
\caption{Flow of an electrically-conducting newtonian fluid subject to a uniform vertical magnetic field $B_0$. It is driven by a horizontal body-force $\textbf{f}$ that is independent of the vertical. We assume periodic boundary-conditions in the horizontal, and stress-free boundaries with large magnetic-permeability  at $z=0$ and $z=H$.}
\label{3Dbox}
\end{figure}

We consider the setup sketched in figure \ref{3Dbox}: an incompressible fluid of kinematic viscosity $\nu$, density $\rho$ and electrical conductivity $\sigma$ flows inside a domain $(x,y,z) \in \mathcal{D}=[0, L] \times [0,L] \times [0,H]$ with a Cartesian frame $(\textbf{e}_x,\textbf{e}_y,\textbf{e}_z)$. It is stirred by a time-independent divergence-free two-dimensional horizontal body-force $\textbf{f}(x,y)=(f_x,f_y,0)$ that is periodic on a scale $\ell$, an integer fraction of $L$. That is, ${\bf f}(x,y)=F {\boldsymbol \phi}(\frac{x}{\ell},\frac{y}{\ell})$, where ${\boldsymbol \phi}$  is periodic of period $1$ in each dimensionless variable and has an rms magnitude $1$. We refer to $F$ as the amplitude, and ${\boldsymbol \phi}$ as the shape of the force. 

A homogeneous steady vertical magnetic field $B_0 \textbf{e}_z$ is applied to the system. We consider periodic boundary conditions in the horizontal directions ($x$ and $y$), and stress-free boundary conditions for the velocity field at the top and bottom boundaries. We further limit attention to the case that these two boundaries have very large magnetic permeability, so that the magnetic field is vertical at $z=0$ and $z=H$. Denoting $\textbf{u}(x,y,z,t)$ and $\textbf{B}(x,y,z,t)$ as the velocity and magnetic fields inside the fluid, the boundary conditions at the top and bottom boundaries of the domain are
\begin{equation}
\left.
\begin{array}{ll}
\partial_z u_x=0\,, \partial_z u_y=0\,, u_z=0 \\
B_x=0\,, B_y=0\,, \partial_z B_z=0
\end{array}
\right\} \mbox{at } z=0 \mbox{ and } z=H \, .
\label{BC}
\end{equation}




The magnetic field inside the fluid follows the induction equation, which using the decomposition $\textbf{B}(x,y,z,t)=B_0 \textbf{e}_z+ \textbf{b}(x,y,z,t)$ reads
\begin{equation}
\partial_t \textbf{b} + (\textbf{u} \cdot \bnabla) \textbf{b} - (\textbf{b} \cdot \bnabla) \textbf{u}= B_0 \partial_z \textbf{u} +\eta \Delta  \textbf{b}\, , \qquad \bnabla.\textbf{b} =0 \label{induction}
\end{equation}
where $\eta=1/(\mu_0 \sigma)$, with $\mu_0$ the magnetic permeability of vacuum. Without loss of generality, we assume $B_0>0$ in the following.

The velocity field follows the body-forced Navier-Stokes equation with the Lorentz force
\begin{equation}
\partial_t \textbf{u} +(\textbf{u} \cdot \bnabla) \textbf{u}=-\bnabla{p} + \nu \Delta \textbf{u} + \textbf{f}(x,y) + \frac{1}{\rho \mu_0} (\bnabla \times \textbf{b} )\times(B_0 \textbf{e}_z+ \textbf{b})\, , \qquad \bnabla.\textbf{u} =0 \, . \label{NS}
\end{equation}


From equations (\ref{induction}) and (\ref{NS}) we define the kinetic and magnetic Reynolds numbers using the root-mean-square velocity $U$, where the mean is performed over space and time:
\begin{equation}
Re=\frac{U \ell}{\nu}\, ,  \qquad Rm=\frac{U \ell}{\eta} \,.
\end{equation}
In the following, we focus on weak magnetic Reynolds numbers, $Rm \lsim 1$, so that the dominant balance in (\ref{induction}) is
\begin{equation}
B_0 \frac{U}{\ell} \sim \frac{\eta b}{\ell^2}\, .
\end{equation}
One can define a third independent dimensionless quantity, the interaction parameter $N$, to evaluate the ratio between the Lorentz force and the inertial term in the Navier-Stokes equation:
\begin{equation}
N=\frac{\sigma B_0^2 \ell}{\rho U}=\frac{ B_0^2 \ell}{\rho \mu_0 \eta U} \, .
\end{equation}
$N$ is a dimensionless measure of the strength of the external field $B_0$.

This magnetohydrodynamic problem admits two-dimensional solutions of the form $\textbf{u}=\textbf{V}(x,y,t)$, $\textbf{b}=0$, where $\textbf{V}(x,y,t)$ satisfies a standard non-magnetic two-dimensional Navier-Stokes equation. Our goal is to prove that, for large enough interaction parameter, such motion is robust with respect to three-dimensional perturbations, i.e., the three-dimensional part of the velocity field decays in the long-time limit, and the system relaxes to a 2D flow.

\subsection{The quasi-static, or vanishing $Rm$ limit \label{QS}}

The quasi-static limit consists in going to the limit $Rm \to 0$ while the interaction parameter $N$ remains of order unity, before expanding the equations to lowest order in $Rm$. At this order, the induction equation (\ref{induction}) reduces to $\eta \Delta {\bf b} = - B_0 \partial_z {\bf u}$, i.e. the induced magnetic field follows the velocity field adiabatically. It is much weaker than the external field, with ${\bf b} \sim Rm B_0$. To lowest order in $Rm$, the Lorentz force is then $(\bnabla \times \textbf{b} )\times B_0 \textbf{e}_z / (\rho \mu_0)$, and the Navier-Stokes equation reads
\begin{equation}
\partial_t \textbf{u} +(\textbf{u} \cdot \bnabla) \textbf{u}=-\bnabla{p} + \nu \Delta \textbf{u} + \textbf{f}(x,y) - \beta \Delta^{-1}  \partial_{zz} {\bf u} \, , \qquad \bnabla.\textbf{u} =0 \, , \label{NSzeroRm}
\end{equation}
where $\Delta^{-1}$ is the inverse Laplacian operator, which can be thought of as a division by $k^2$ of each Fourier component, $k$ denoting the wavenumber. $\beta$ is the inverse Joule time scale corresponding to magnetic damping, $\beta=\sigma B_0^2/\rho =B_0^2/(\rho \mu_0 \eta)$. It is related to the interaction parameter through $N=\beta \ell / U$. The main convenience of the quasi-static equation (\ref{NSzeroRm}) is that it involves the velocity field only.

\subsection{The energy dissipation of 2D and 3D turbulence}

As mentioned in the introduction, the central challenge of the analysis is to establish the stability of some vertically-invariant flow ${\bf V}(x,y,t)$ with respect to vertically-dependent perturbations ${\bf v}(x,y,z,t)$, without knowing the details of the base flow ${\bf V}$: it is typically a turbulent solution of a two-dimensional body-forced Navier-Stokes system. Instead of a precise knowledge of ${\bf V}(x,y,t)$, for the proofs we make use of rigorous bounds on the time-averaged energy dissipation for solutions of the body-forced 2D and 3D Navier-Stokes equations. The corresponding bounds on the gradients of ${\bf V}$ are sufficient to establish its stability with respect to 3D perturbations. We now recall these bounds, committing the proofs to an appendix.

For hydrodynamic turbulence in a 3D periodic domain, \citet{Doering} derived a bound on the time-averaged viscous energy dissipation of any solution ${\bf u}$ to the body-forced Navier-Stokes equation, i.e., equation (\ref{NS}) without the Lorentz force. The method consists in bounding the amplitude of the body-force, and the power it inputs, in terms of the root-mean-square velocity $U$. We prove in appendix \ref{appbound} that the bound easily carries over to the present magnetohydrodynamic system in the quasi-static limit, i.e., when it is governed by (\ref{NSzeroRm}). In the following, the  constants with subscripts $(c_i)_{i\in\mathbb{N}}$ are dimensionless positive numbers that depend on the {\it shape} of the forcing function only, not on its amplitude or length scale $\ell$. By contrast, the undecorated $c$ denotes the generic positive $O(1)$ constant that does not depend on any parameter of the problem. With these notations, the bound reads

\begin{equation}
\left<  \| \bnabla \textbf{u}  \|_{2}^2 \right>   \leq \frac{U^3 L^2 H}{\ell \nu} \left( \frac{c_1}{Re} +c_2 \right) \, ,
\label{DF}
\end{equation}
where $\| \dots \|_2^2$ is the standard $L_2$ norm over $\mathcal{D}$, $\left< \dots \right>$ denotes time-average, and $c_1$ and $c_2$ depend only on the shape of the forcing function ${\bf f}$.

Bounds can also be obtained for solutions of the 2D Navier-Stokes equation, i.e., for vertically-invariant horizontal velocity fields satisfying equation (\ref{NS}). Denoting the vorticity as $\omega$, \citet{Alexakis} derived a bound on time-averaged dissipation rate of enstrophy $\| \omega \|_2^2$ in 2D. For a time-independent forcing function, and using our notations, their equation (19) translates into
\begin{eqnarray}
 \left< \| \bnabla \omega \|_2^2 \right> \leq \frac{H L^2 U^3}{\nu \ell^3} \left( c_3 + \frac{c_4}{Re} \right) \, , \label{2Dbound}\\
\end{eqnarray}
where $c_3$ and $c_4$ are again dimensionless constants that depend only on the shape of ${\bf f}$, and the prefactor $H$ arises from $\| \dots \|_2$ being the $L_2$ norm in 3D.

This bound for vertically-invariant flows can be further reduced if the forcing is such that $\bnabla^2 {\bf f}$ is proportional to ${\bf f}$. These harmonic forcings are sometimes called ``single-mode", or Kolmogorov forcings. For such forcings an improved bound on the enstrophy dissipation rate is
\begin{equation}
\left< \| \bnabla \omega \|_2^2 \right> \leq c \frac{U^2 L^2 H}{\ell^4} \, . \label{2DboundKolmogorov}
\end{equation}


\section{Absolute two-dimensionalization for vanishing magnetic Reynolds number\label{section3}}


\subsection{Evolution equation for the kinetic energy of the three-dimensional part of the velocity field\label{StratAbs}}

We consider the evolution of an initially 3D velocity field under equation (\ref{NSzeroRm}). Define the vertical average of a field ${\bf h}$ as
\begin{equation}
\bar{{\bf h}}(x,y,t) = \frac{1}{H} \int_{z=0}^{z=H}  {\bf h}(x,y,z,t) \, \mathrm{d}z \, ,
\end{equation}
and decompose the velocity field into 
\begin{equation}
\textbf{u}=\textbf{V}(x,y,t)+\textbf{v}(x,y,z,t) \mbox{, with } \textbf{V}=\bar{\textbf{u}} \mbox{, and } \textbf{v}=\textbf{u}-\textbf{V} \, .
\label{decomp}
\end{equation}
In the following we call ${\bf V}$ (resp. ${\bf v}$) the two-dimensional (resp. three-dimensional) part of the velocity field. Our goal is to find a criterion that insures the decay of ${\bf v}$. One can obtain the evolution equation for the two-dimensional part of the velocity field by vertically-averaging the Navier-Stokes equation (\ref{NSzeroRm}),
\begin{equation}
\partial_t \textbf{V} +(\textbf{V} \cdot \bnabla) \textbf{V} +\overline{(\textbf{v} \cdot \bnabla) \textbf{v}}=-\bnabla{\bar{p}} + \nu \Delta \textbf{V} + \textbf{f}(x,y) \, .
\label{eqV}
\end{equation}
By subtracting this equation to the Navier-Stokes equation, we get the evolution equation for the three-dimensional part of the flow:
\begin{equation}
\partial_t \textbf{v} +(\textbf{V} \cdot \bnabla) \textbf{v} +(\textbf{v} \cdot \bnabla) \textbf{V} +(\textbf{v} \cdot \bnabla) \textbf{v} - \overline{(\textbf{v} \cdot \bnabla) \textbf{v}}=-\bnabla(p-\bar{p}) + \nu \Delta \textbf{v}  - \beta \Delta^{-1}  \partial_{zz} \textbf{v}  \, .
\label{eqv}
\end{equation}
The incompressibility constraint imposes $\bnabla \cdot \textbf{V} = 0$ and $\bnabla \cdot \textbf{v} = 0$. To get the evolution equation for the kinetic energy of the 3D part of the velocity field, one takes the dot product of (\ref{eqv}) with $\textbf{v}$ and integrates over the domain. Some integrations by parts using the incompressibility constraint lead to
\begin{equation}
\frac{\mathrm{d}}{\mathrm{d}t} \left( \frac{1}{2} \|\textbf{v}\|_2^2 \right) = \mathcal{H}\{\textbf{v},t\}  ,
\label{eqv2}
\end{equation}
with $\mathcal{H}$ a time-dependent quadratic functional of the field $\textbf{v}$,
\begin{equation}
\mathcal{H}\{\textbf{v},t\} = - \int_\mathcal{D} \textbf{v} \cdot (\bnabla \textbf{V}) \cdot \textbf{v} \, \mathrm{d}^3 \textbf{x} - \nu \|\bnabla \textbf{v}\|_2^2 - \beta \|\bnabla^{-1}  \partial_z{\textbf{v}} \|_2^2\, ,
\label{defH}
\end{equation}
where we introduced the notation $\bnabla^{-1}=\bnabla \Delta^{-1}$.

The strategy is the following: we seek a function $\lambda_1(t)$ such that
\begin{itemize}
\item For all divergence-free fields $\textbf{v}$ with zero vertical average ($\bar{\textbf{v}}=\textbf{0}$) satisfying the boundary conditions (\ref{BC}),
\begin{equation}
\mathcal{H}\{\textbf{v},t\} \leq \lambda_1(t) \| \textbf{v} \|_2^2 .
\label{ineqlambda}
\end{equation}
\item The long-time average of $\lambda_1$ is negative, i.e. $\left< \lambda_1 \right> <0$.
\end{itemize}

Then the kinetic energy in the three-dimensional part of the velocity field decays at least exponentially in time. The velocity field then becomes two-dimensional asymptotically in the long-time limit, regardless of the amplitude of the initial three-dimensional perturbation ${\bf v}$. In the following, we call this property of the system{ \it absolute two-dimensionalization}. Indeed, our approach for proving two-dimensionalization is very similar to proving absolute stability of some velocity field $\textbf{V}$ to perturbations $\textbf{v}$. However, it is conceptually distinct from a standard absolute stability analysis: in relation (\ref{defH}), $\textbf{V}(x,y,t)$ is a possibly turbulent velocity field that satisfies equation (\ref{eqV}), which includes an undetermined Reynolds stress term.

\subsection{A sufficient condition for two-dimensionalization}

To find a function $\lambda_1$ that satisfies (\ref{ineqlambda}), we need to bound the first term on the right-hand-side of (\ref{defH}). This task is performed in appendix \ref{appendix1}, where we use Ladyzhenskaya's inequality on the $L_4$ norm.  We obtain 
\begin{eqnarray}
 \left| \int_\mathcal{D} \textbf{v} \cdot (\bnabla \textbf{V}) \cdot \textbf{v} \, \mathrm{d}^3 \textbf{x} \right| & \leq &  c \left( \frac{ \| \bnabla \textbf{V} \|_2}{L \sqrt{H}} + \frac{\| \bnabla \textbf{V} \|_2^2}{\nu H}  \right) \| {\bf v} \|_2^2 + \frac{\nu}{2}  \| \bnabla {\bf v} \|_2^2 \, .
\end{eqnarray}
We bound the last term in (\ref{defH}) using Poincar\'e's inequality in the $z$ direction only,
\begin{equation}
- \beta \|\bnabla^{-1}  \partial_z{\textbf{v}} \|_2^2 = - \beta \| \partial_z \bnabla^{-1}  {\textbf{v}} \|_2^2 \leq - \beta \frac{\pi^2}{H^2} \| \bnabla^{-1}  {\textbf{v}} \|_2^2 \, ,
\end{equation}
to get the following bound on the functional $\mathcal{H}$,
\begin{equation}
\mathcal{H}\{\textbf{v},t\} \leq c \left( \frac{ \| \bnabla \textbf{V} \|_2}{L \sqrt{H}} + \frac{\| \bnabla \textbf{V} \|_2^2}{\nu H}  \right) \| {\bf v} \|_2^2 - \beta \frac{\pi^2}{H^2} \| \bnabla^{-1}  {\textbf{v}} \|_2^2 - \frac{\nu}{2}  \| \bnabla {\bf v} \|_2^2 \, . \label{boundH}
\end{equation}
At this stage it is convenient to decompose $\textbf{v}$ into Fourier components,
\begin{equation}
\textbf{v} = \sum\limits_{\textbf{k}} \hat{\textbf{v}}_\textbf{k}(t) e^{i \textbf{k}\cdot \textbf{x}} + c.c. \, ,
\label{Fourier3D}
\end{equation}
where the wavevector $\textbf{k}$ takes the values 
\begin{equation}
\textbf{k}=(k_x,k_y,k_z)=\left(n_x \frac{2\pi}{L}, n_y \frac{2\pi}{L},n_z  \frac{\pi}{H} \right), \mbox{with } {(n_x,n_y,n_z) \in \mathbb{N}\times\mathbb{N} \times\mathbb{N}^*}. 
\end{equation}
The value $n_z=0$ is excluded because $\textbf{v}$ is the three-dimensional part of the velocity field and therefore it has zero vertical average.

Using this Fourier decomposition together with Parseval's equality, equation (\ref{boundH}) gives
\begin{eqnarray}
\nonumber \mathcal{H}\{\textbf{v},t\}  & \leq &c \left( \frac{ \| \bnabla \textbf{V} \|_2}{L \sqrt{H}} + \frac{\| \bnabla \textbf{V} \|_2^2}{\nu H}  \right) \| {\bf v} \|_2^2 - 2L^2 H \sum\limits_{\textbf{k}} \left(    \frac{\beta \pi^2}{k^2 H^2} + \frac{\nu}{2} k^2 \right)    | \hat{\textbf{v}}_\textbf{k}|^2 \\
 & \leq & \left[ c \left( \frac{ \| \bnabla \textbf{V} \|_2}{L \sqrt{H}} + \frac{\| \bnabla \textbf{V} \|_2^2}{\nu H}  \right) -\min_{k \in \mathbb{R}^+} \left(    \frac{\beta \pi^2}{k^2 H^2} + \frac{\nu}{2} k^2 \right) \right] \| {\bf v} \|_2^2 \, ,\label{Fourierineq}
\end{eqnarray}
which we write as
\begin{eqnarray}
\mathcal{H}\{\textbf{v},t\} & \leq &  \lambda_1(t) \| \textbf{v} \|_{2}^2 \mbox{ , with } \lambda_1(t)=  c \left( \frac{ \| \bnabla \textbf{V} \|_2}{L \sqrt{H}} + \frac{\| \bnabla \textbf{V} \|_2^2}{\nu H}  \right) - \frac{\pi}{H} \sqrt{2 \beta \nu} \, .
\end{eqnarray}

To prove the decay of the three-dimensional part of the velocity field in the long-time limit, we need the time-average of $\lambda_1$ to be negative. Using the concavity of the square-root function, we have $\la \| \bnabla \textbf{V} \|_2 \ra \leq \sqrt{ \la \| \bnabla \textbf{V} \|_2^2 \ra}$, hence the following upper bound on $\la \lambda_1 \ra$,
\begin{equation}
\la \lambda_1 \ra \leq c \left( \frac{1 }{L \sqrt{H}} \sqrt{ \la \| \bnabla \textbf{V} \|_2^2 \ra } + \frac{1}{\nu H} \la \| \bnabla \textbf{V} \|_2^2 \ra \right) - \frac{\pi}{H} \sqrt{2 \beta \nu} \, .
\end{equation}
If the right-hand side expression is negative, then so is $\la \lambda_1 \ra$ and the flow becomes exactly 2D in the long-time limit. This expression is a quadratic polynomial in $\sqrt{ \la \| \bnabla \textbf{V} \|_2^2 \ra }$. From the computation of the roots of the polynomial, we obtain that it is negative if
\begin{equation}
\la \| \bnabla \textbf{V} \|_2^2 \ra  \leq c \sqrt{\beta} \nu^{3/2} \, ,
\label{criterionAbs}
\end{equation}
where $c$ is yet another dimensionless positive constant that is independent of the parameters of the problem.

%
The term $\left<  \| \bnabla \textbf{V}  \|_{2}^2 \right> $ can be bounded in terms of the viscous energy dissipation per unit mass $\epsilon$ inside the flow,
\begin{equation}
\left<  \| \bnabla \textbf{V}  \|_{2}^2 \right>  \leq \left<  \| \bnabla \textbf{u}  \|_{2}^2 \right>  = \frac{\epsilon L^2 H}{\nu} \leq \frac{U^3 L^2 H}{\ell \nu} \left( \frac{c_1}{Re} +c_2 \right) \, ,
\label{dV2}
\end{equation}
where we used the bound (\ref{DF}) on the energy dissipation of a 3D flow. 
We obtain a sufficient criterion for two-dimensionalization expressed in terms of the root-mean-square velocity by demanding that the right-hand side of (\ref{criterionAbs}) be greater than the right-hand-side of (\ref{dV2}):
\begin{equation}
N \geq c\, \frac{ H^2 L^4}{ \ell^6} Re^5 \left( \frac{c_1}{Re} +c_2 \right)^2 \, .
\label{NcAbsolute}
\end{equation}

This criterion shows unambiguously that, whatever the Reynolds number is, there is a critical value of the interaction parameter above which the three-dimensional part of the velocity field decays and the flow becomes two-dimensional in the long-time limit. This two-dimensionalization occurs for three-dimensional perturbations ${\bf v}$ of arbitrary amplitude, provided this initial condition leads an rms velocity satisfying (\ref{NcAbsolute}), and we therefore call it {\it absolute two-dimensionalization}. 

As a side note, we stress the difference between specifying the rms velocity $U$ and specifying the amplitude of the forcing. All the initial conditions leading to an rms velocity $U$ satisfying (\ref{NcAbsolute}) two-dimensionalize in the long-time limit. By contrast, for constant amplitude of the forcing, two different initial conditions may end up in different attractors of the system, with differing rms velocities $U_1$ and $U_2$. It may be that $U_1$ satisfies (\ref{NcAbsolute}) but $U_2$ does not.


\section{Linear two-dimensionalization\label{section4}}

In this section we come back to the full-MHD problem, with the goal of proving two-dimensionalization for finite values of $Rm$. We show that, starting from an infinitesimal three-dimensional velocity perturbation, the flow becomes two-dimensional asymptotically in time provided the interaction parameter is above some threshold value and the magnetic Reynolds number is lower than some other threshold value.

Let us decompose the velocity field into $\textbf{u}=\textbf{V}(x,y,t)+\textbf{v}(x,y,z,t)$, where $\textbf{V}(x,y,t)$ is a solution of the 2D body-forced Navier-Stokes equation, and assume that $\textbf{v}(x,y,z,t)$ and $\textbf{b}(x,y,z,t)$ are infinitesimal perturbations to this (possibly turbulent) base state. The three fields $\textbf{V}$, $\textbf{v}$ and $\textbf{b}$ are divergence-free. The linearized evolution equations for $\textbf{v}$ and $\textbf{b}$ are
\begin{eqnarray}
& & \partial_t \textbf{v} +(\textbf{v} \cdot \bnabla) \textbf{V} + (\textbf{V} \cdot \bnabla) \textbf{v} =-\bnabla{p} + \nu \Delta \textbf{v} +   \frac{1}{\rho \mu_0} B_0 \partial_z \textbf{b}\, , \label{linNS} \\
& & \partial_t \textbf{b} + (\textbf{V} \cdot \bnabla) \textbf{b} = B_0 \partial_z \textbf{v} +(\textbf{b} \cdot \bnabla) \textbf{V} +\eta \Delta  \textbf{b}\, , \label{lininduction}
\end{eqnarray}
where $p$ is the pressure perturbation.
From these equations, we compute evolution equations for the three following quadratic quantities
\begin{eqnarray}
\frac{\mathrm{d}}{\mathrm{d}t} \left( \frac{1}{2} \|\textbf{v}\|_2^2 \right) & = &  - \int_\mathcal{D} \textbf{v} \cdot (\bnabla \textbf{V}) \cdot \textbf{v} \, \mathrm{d}^3 \textbf{x} - \nu \|\bnabla \textbf{v}\|_2^2 + \frac{B_0}{\rho \mu_0} \int_\mathcal{D} \textbf{v} \cdot \partial_z \textbf{b} \, \mathrm{d}^3 \textbf{x}  \label{eqken} \, ,\\
\frac{\mathrm{d}}{\mathrm{d}t} \left( \frac{1}{2} \|\textbf{b}\|_2^2 \right) & = &   \int_\mathcal{D} \textbf{b} \cdot (\bnabla \textbf{V}) \cdot \textbf{b} \, \mathrm{d}^3 \textbf{x} - \eta \|\bnabla \textbf{b}\|_2^2 - \frac{B_0}{\rho \mu_0} \int_\mathcal{D} \textbf{v} \cdot \partial_z \textbf{b} \, \mathrm{d}^3 \textbf{x}  \label{eqmen} \, , \\
\frac{\mathrm{d}}{\mathrm{d}t} \left(  \int_\mathcal{D} \textbf{v} \cdot \partial_z \textbf{b} \, \mathrm{d}^3 \textbf{x} \right) & = &   \int_\mathcal{D} \partial_z\textbf{v} \cdot (\bnabla \textbf{V}) \cdot \textbf{b} \, \mathrm{d}^3 \textbf{x}  -   \int_\mathcal{D} \textbf{b} \cdot (\bnabla \textbf{V}) \cdot \partial_z \textbf{v} \, \mathrm{d}^3 \textbf{x}       \label{eqcross} \\
\nonumber & & +(\nu+ \eta)  \int_\mathcal{D} (\bnabla \partial_z\textbf{v}) \cdot (\bnabla \textbf{b}) \, \mathrm{d}^3 \textbf{x}  + B_0 \left(\frac{1}{\rho \mu_0} \|\partial_z \textbf{b}\|_2^2 - \|\partial_z \textbf{v}\|_2^2 \right) \, .
\end{eqnarray}

With the goal of determining a Lyapunov functional for the system of linearized equations, let us define 
\begin{equation}
\mathcal{F}_\alpha\{\textbf{v},\textbf{b} \}=\frac{\|\textbf{v}\|_2^2 }{2}+    \frac{\|\textbf{b}\|_2^2}{2 \rho \mu_0} + \alpha   \int_\mathcal{D} \textbf{v} \cdot \partial_z \textbf{b} \, \mathrm{d}^3 \textbf{x}\, , 
\end{equation}
where $\alpha$ is a positive real number, and combine (\ref{eqken}), (\ref{eqmen}) and (\ref{eqcross}) to obtain
\begin{eqnarray}
\label{ineq1} \frac{\mathrm{d}}{\mathrm{d}t} \mathcal{F}_\alpha & = &  - \int_\mathcal{D} \textbf{v} \cdot (\bnabla \textbf{V}) \cdot \textbf{v} \, \mathrm{d}^3 \textbf{x} - \nu \|\bnabla \textbf{v}\|_2^2   + \frac{1}{\rho \mu_0}\int_\mathcal{D} \textbf{b} \cdot (\bnabla \textbf{V}) \cdot \textbf{b} \, \mathrm{d}^3 \textbf{x}  \\
\nonumber & & - \frac{\eta}{\rho \mu_0} \|\bnabla \textbf{b}\|_2^2+ \alpha \left[   \int_\mathcal{D} \partial_z\textbf{v} \cdot (\bnabla \textbf{V}) \cdot \textbf{b} \, \mathrm{d}^3 \textbf{x}  -   \int_\mathcal{D} \textbf{b} \cdot (\bnabla \textbf{V}) \cdot \partial_z \textbf{v} \, \mathrm{d}^3 \textbf{x}  \right.  \\
\nonumber & & \left.  +(\nu+ \eta)  \int_\mathcal{D} (\bnabla \partial_z\textbf{v}) \cdot (\bnabla \textbf{b}) \, \mathrm{d}^3 \textbf{x}  + B_0 \left(\frac{1}{\rho \mu_0} \|\partial_z \textbf{b}\|_2^2 - \|\partial_z \textbf{v}\|_2^2 \right)   \right] \, .
\end{eqnarray}

Because the base state is independent of the vertical and equations (\ref{linNS}) and (\ref{lininduction})  are linear, we can decompose the perturbation into Fourier modes in $z$ and study the independent evolution of a single vertical Fourier mode. In the following, we therefore consider that $(\textbf{v},\textbf{b})$ is a single vertical Fourier mode of perturbation, with vertical wavenumber $q$. Then $\|\partial_z \textbf{v}\|_2^2=q^2 \| \textbf{v}\|_2^2$ and $\|\partial_z \textbf{b}\|_2^2=q^2 \| \textbf{b}\|_2^2$. We denote as $\mathcal{F}^{(q)}_{\alpha_q}$ the functional $\mathcal{F}_\alpha$ for such a Fourier mode, where the value of $\alpha_q$ is free to depend on $q$.  Using H\"older's, the Cauchy-Schwarz and Young's inequalities, we obtain from (\ref{ineq1})
\begin{eqnarray}
\label{ineq2}   \frac{\mathrm{d}}{\mathrm{d}t}    \mathcal{F}^{(q)}_{\alpha_q}    & \leq &   \|\bnabla \textbf{V}\|_\infty \|\textbf{v}\|_2^2 - \nu \|\bnabla \textbf{v}\|_2^2 + \|\bnabla \textbf{V}\|_\infty \frac{\|\textbf{b}\|_2^2}{\rho \mu_0} -\frac{\eta}{\rho \mu_0} \|\bnabla \textbf{b}\|_2^2 \\
\nonumber & + & {\alpha_q} \left[ \|\bnabla \textbf{V}\|_\infty \left( \frac{1}{{\alpha_q}} \|\textbf{v}\|_2^2 + {\alpha_q} q^2 \|\textbf{b}\|_2^2  \right) \right.     \\
\nonumber & + & \frac{1+Pm}{\sqrt{Pm}} \sqrt{\rho \mu_0} \frac{H}{\pi} \left( \frac{\nu}{2} q^2 \|\bnabla \textbf{v}\|_2^2   + \frac{\eta \pi^2}{2 \rho \mu_0 H^2} \|\bnabla \textbf{b}\|_2^2    \right) \\
\nonumber      & + & \left. B_0 q^2\left(\frac{1}{\rho \mu_0} \| \textbf{b}\|_2^2 - \| \textbf{v}\|_2^2 \right)   \right] \, ,
\end{eqnarray}
where we used Young's inequality (together with an optimization) to get the terms in the parentheses of the second and third lines, and we introduced the magnetic Prandtl number $Pm=\nu/\eta$ to alleviate somewhat the equations.

We then decompose the Ohmic dissipation term into 
\begin{equation}
-\frac{\eta}{\rho \mu_0} \|\bnabla \textbf{b}\|_2^2 \leq -\frac{\eta}{2 \rho \mu_0} \|\bnabla \textbf{b}\|_2^2 -\frac{q^2 \eta}{2\rho \mu_0} \| \textbf{b}\|_2^2 \, ,
\end{equation}
and rewrite (\ref{ineq2}) as
\begin{eqnarray}
 \frac{\mathrm{d}}{\mathrm{d}t}  \mathcal{F}^{(q)}_{\alpha_q}   & \leq & \| \textbf{v}\|_2^2 \left[ 2 \|\bnabla \textbf{V}\|_\infty  - {\alpha_q} B_0 q^2 \right] + \nu \| \bnabla \textbf{v}\|_2^2 \left[ -1 + \frac{{\alpha_q} q^2  H \sqrt{\rho \mu_0} (1+Pm)}{2 \pi \sqrt{Pm}}  \right] \label{4brackets} \\
\nonumber & + & \frac{\| \textbf{b}\|_2^2}{\rho \mu_0} \left[  \|\bnabla \textbf{V}\|_\infty (1+{\alpha_q}^2 q^2 \rho \mu_0)  + {\alpha_q} B_0 q^2  -\frac{\eta q^2}{2}\right]  \\
\nonumber &+& \eta \frac{\|\bnabla \textbf{b}\|_2^2 }{\rho \mu_0} \left[ -\frac{1}{2} + \frac{{\alpha_q} \pi^2 \sqrt{\rho \mu_0} (1+Pm)}{2 H^2 \sqrt{Pm}}    \right] \, .
\end{eqnarray}
The method is now similar to the one used in section \ref{StratAbs}: if we can find a value of $\alpha_q$ such that
\begin{itemize}
\item $\mathcal{F}^{(q)}_{\alpha_q}$ is a positive functional,
\item there exists a real function $\lambda_2(t)$ such that $\frac{\mathrm{d}}{\mathrm{d}t} \mathcal{F}^{(q)}_{\alpha_q} \leq \lambda_2(t) \mathcal{F}^{(q)}_{\alpha_q}$,
\item the time average of $\lambda_2(t)$ is strictly negative: $\left< \lambda_2(t) \right> < 0$,
\end{itemize}
then the vertical Fourier mode $q$ of the perturbation decays to zero in the long-time limit. And if we can find such an $\alpha_q$ for all Fourier modes with nonzero vertical wave number $q$ allowed in the system, then the $L_2$ norm of the $z$-dependent perturbation decays to zero in the long-time limit, and the system relaxes to the 2D solution of the Navier-Stokes equation.

To find a suitable $\alpha_q$, we first impose the constraint
\begin{equation}
0 <  \alpha_q < \frac{\pi \sqrt{Pm}}{q^2 H \sqrt{\rho \mu_0} (1+Pm)}\, . \label{C1}
\end{equation}
The reason for this constraint is made clear before equation (\ref{drop}): it ensures that the terms proportional to $\|\bnabla \textbf{v}\|_2^2$ and $\|\bnabla \textbf{b}\|_2^2$ in (\ref{4brackets}) are negative and can be discarded.

This constraint, together with $q \geq \pi/H$, Cauchy-Schwarz and Young's inequalities, leads to
\begin{eqnarray}
\left| \alpha_q  \int_\mathcal{D} \textbf{v} \cdot \partial_z \textbf{b} \, \mathrm{d}^3 \textbf{x} \right| & \leq & \alpha_q |q| \| \textbf{v}\|_2 \| \textbf{b}\|_2 \\
\nonumber & \leq & \alpha_q |q| \sqrt{\rho \mu_0} \left( \frac{\| \textbf{v}\|_2^2}{2} + \frac{\| \textbf{b}\|_2^2}{2 \rho \mu_0} \right) \\
\nonumber & \leq & \frac{\sqrt{Pm}}{1+Pm} \left( \frac{\| \textbf{v}\|_2^2}{2} + \frac{\| \textbf{b}\|_2^2}{2 \rho \mu_0} \right)\, ,
\end{eqnarray}
so that an upper and a lower bound on the functional $\mathcal{F}^{(q)}_{\alpha_q}$ are
\begin{equation}
 0 \leq \left( \frac{\| \textbf{v}\|_2^2}{2} + \frac{\| \textbf{b}\|_2^2}{2 \rho \mu_0} \right) \left( 1-\frac{\sqrt{Pm}}{1+Pm} \right)  \leq \mathcal{F}^{(q)}_{\alpha_q}  \leq \left( \frac{\| \textbf{v}\|_2^2}{2} + \frac{\| \textbf{b}\|_2^2}{2 \rho \mu_0} \right) \left( 1+\frac{\sqrt{Pm}}{1+Pm} \right)  \, , 
\end{equation}
or, since we do not care about optimizing the prefactors,
\begin{equation}
0  \leq  \frac{1}{2} \left( \frac{\| \textbf{v}\|_2^2}{2} + \frac{\| \textbf{b}\|_2^2}{2 \rho \mu_0} \right) \leq \mathcal{F}^{(q)}_{\alpha_q} \leq \frac{3}{2} \left( \frac{\| \textbf{v}\|_2^2}{2} + \frac{\| \textbf{b}\|_2^2}{2 \rho \mu_0} \right)    \, , \label{boundF}
\end{equation}
where we used the fact that $\frac{\sqrt{Pm}}{1+Pm}<\frac{1}{2}$. Inequality (\ref{boundF}) proves that $\mathcal{F}^{(q)}_{\alpha_q}$ is a positive functional. 

From the constraint (\ref{C1}) and using again $q \geq \pi/H$, we show that the terms proportional to $\|\bnabla \textbf{v}\|_2^2$ and $\|\bnabla \textbf{b}\|_2^2$ in (\ref{4brackets}) are negative, and upon dropping them we get
\begin{eqnarray}
 \frac{\mathrm{d}}{\mathrm{d}t}  \mathcal{F}^{(q)}_{\alpha_q}   & \leq & \| \textbf{v}\|_2^2 \left[ 2 \|\bnabla \textbf{V}\|_\infty  - \alpha_q B_0 q^2 \right]  \label{drop}\\
 \nonumber & & +  \frac{\| \textbf{b}\|_2^2}{\rho \mu_0} \left[  \|\bnabla \textbf{V}\|_\infty (1+\alpha_q^2 q^2 \rho \mu_0)  + \alpha_q B_0 q^2  -\frac{\eta q^2}{2}\right]  \\
 \nonumber & \leq & 4 \|\bnabla  \textbf{V}\|_\infty \left( \frac{\| \textbf{v}\|_2^2}{2} + \frac{\| \textbf{b}\|_2^2}{2 \rho \mu_0} \right) -\alpha_q B_0 q^2 \| \textbf{v}\|_2^2 + \left( \alpha_q B_0 q^2 -\frac{\eta \pi^2}{2H^2} \right) \frac{\| \textbf{b}\|_2^2}{\rho \mu_0}
\end{eqnarray}
Let us finally pick 
\begin{equation}
\alpha_q=\frac{7}{B_0 q^2} \left< \|\bnabla  \textbf{V}\|_\infty \right> \label{valalpha} \, .
\end{equation}
(remember that we consider $B_0>0$). We can do so if this value of $\alpha_q$ satisfies the constraint (\ref{C1}). Then
\begin{eqnarray}
 \frac{\mathrm{d}}{\mathrm{d}t}  \mathcal{F}^{(q)}_{\alpha_q}  & \leq & 4 \|\bnabla  \textbf{V}\|_\infty \left( \frac{\| \textbf{v}\|_2^2}{2} + \frac{\| \textbf{b}\|_2^2}{2 \rho \mu_0} \right)-14 \left< \|\bnabla  \textbf{V}\|_\infty \right> \frac{\| \textbf{v}\|_2^2}{2} \\
 \nonumber & & +\left(14 \left< \|\bnabla  \textbf{V}\|_\infty \right> -\frac{\eta \pi^2}{H^2} \right) \frac{\| \textbf{b}\|_2^2}{2\rho \mu_0}\, .
\end{eqnarray}
Let us further assume that we are in the region of parameter space where
\begin{equation}
\frac{\eta \pi^2}{H^2} \geq 28 \left< \|\bnabla  \textbf{V}\|_\infty \right>, \label{C2}
\end{equation}
so that
\begin{eqnarray}
 \frac{\mathrm{d}}{\mathrm{d}t}  \mathcal{F}^{(q)}_{\alpha_q}  & \leq & 4 \|\bnabla  \textbf{V}\|_\infty \left( \frac{\| \textbf{v}\|_2^2}{2} + \frac{\| \textbf{b}\|_2^2}{2 \rho \mu_0} \right)-14 \left< \|\bnabla  \textbf{V}\|_\infty\right> \left( \frac{\| \textbf{v}\|_2^2}{2} + \frac{\| \textbf{b}\|_2^2}{2 \rho \mu_0} \right)  \, .
\end{eqnarray}
Using respectively the lower and upper bounds (\ref{boundF}) on $\mathcal{F}^{(q)}_{\alpha_q}$ to bound the positive and negative terms in the right-hand-side of this inequality, we finally obtain
\begin{equation}
 \frac{\mathrm{d}}{\mathrm{d}t}  \mathcal{F}^{(q)}_{\alpha_q}    \leq   \lambda_2(t) \mathcal{F}^{(q)}_{\alpha_q} \text{  , with  } \lambda_2(t)  =  \left( 8 \|\bnabla  \textbf{V}\|_\infty  -\frac{28}{3} \left< \|\bnabla  \textbf{V}\|_\infty \right> \right)\, \label{eqlambda}
\end{equation}
Because the function $\lambda_2(t)$ has a negative time-average, the three-dimensional perturbation decays in the long-time limit. 
Of course, along the way to obtain (\ref{eqlambda}) we made some assumptions that constrain the region of parameter space in which the conditions hold. Indeed, we need the chosen value (\ref{valalpha}) of $\alpha_q$ to satisfy the constraint (\ref{C1}), and the magnetic diffusivity to be large enough for (\ref{C2}) to hold, i.e., we need respectively
\begin{equation}
\frac{B_0}{\left< \|\bnabla  \textbf{V}\|_\infty \right> H \sqrt{\rho \mu_0}} >  \frac{7(1+Pm)}{\pi \sqrt{Pm}} \text{ , and } \frac{ \left<  \|\bnabla  \textbf{V}\|_\infty \right> H^2}{\eta} \leq \frac{\pi^2}{28} \, . \label{criterion}
\end{equation}
They involve some kind of interaction parameter and magnetic Reynolds number built with the $L_\infty$ norm of the velocity gradients. The former must be large enough and the latter small enough for magnetic damping to be efficient. Because one has little control over the infinite norm of the velocity gradients, we wish to express these constraints using the rms velocity only.


\section{The large Reynolds number regime\label{section5}}

We focus on the large-$Re$ regime and therefore consider $Re \geq 1$. Then, for a solution ${\bf V}(x,y,t)$ of the body-forced two-dimensional Navier-Stokes equation, we can bound the time-averaged $L_\infty$ norm of the velocity gradients $\left< \|\bnabla  \textbf{V}\|_\infty \right>$ in terms of the enstrophy dissipation rate. This task is performed in appendix \ref{Appendix2}, with the following result,
\begin{eqnarray}
\left< \|\bnabla  \textbf{V}\|_\infty \right> & \leq &  c_5 \sqrt{ \frac{\left< \| \bnabla \omega \|_2^2 \right>}{H}} \left( 1+ \ln Re + \ln \frac{L}{\ell}  \right) \, ,
 \label{bndLinfty}
\end{eqnarray}
where the $H$ at the denominator arises from the fact that $\| \dots \|_2$ is $L_2$ norm in 3D, and $c_5$ depends on the shape of the forcing only. The bounds (\ref{2Dbound}) and (\ref{2DboundKolmogorov}) on $\left< \| \bnabla \omega \|_2^2 \right>$ then allow to bound $\left< \|\bnabla  \textbf{V}\|_\infty \right>$ in terms of the rms velocity $U$.

\subsection{Time-independent forcing}

For a time-independent forcing, let us combine the upper bound on the vorticity gradients (\ref{2Dbound}) with equation (\ref{bndLinfty}) to get
\begin{eqnarray}
\left< \|\bnabla  \textbf{V}\|_\infty \right>  & \leq & c_6 \frac{U L}{\ell^2}  \sqrt{Re} \,  \left( 1+ \ln Re + \ln \frac{L}{\ell}  \right) \, ,
\end{eqnarray}
which we substitute into the inequalities (\ref{criterion}) to get alternate sufficient criteria for two-dimensionalization, which now involve only the rms velocity of the base flow:
\begin{eqnarray}
N  & \geq &  c_{7} \frac{H^2 L^2}{\ell^4} Re^2 \left( 1 +  \ln Re +  \ln  \frac{L}{\ell}   \right)^2 \, , \label{condition1}\\
\text{and } Rm & \leq &  \frac{\ell^3}{H^2 L} \, \frac{c_{8}}{\sqrt{Re} \left( 1 +  \ln Re +  \ln  \frac{L}{\ell}   \right) } \, , \label{condition2}
\end{eqnarray}
where we also assumed $Re \geq 1$ and $Pm \leq 1$ along the way, and $c_{7}$ and $c_{8}$ are positive numbers that depend on the forcing shape only. Forgetting about the logarithmic terms, the first condition requires the magnetic field to be strong enough for the interaction parameter to be larger than $Re^2$. Note that the prefactor is proportional to $H^2$, so that the constraint is more easily met in shallow domains. The second condition is to have a low enough magnetic Reynolds number, lower than roughly the inverse square-root of the Reynolds number. Once again, because the prefactor is inversely proportional to $H^2$ this constraint is more easily satisfied in shallow domains. Condition (\ref{condition2}) can be recast in terms of the kinetic Reynolds number using $Rm=Pm \, Re$, which gives
\begin{equation}
Re \leq  \frac{\ell^2}{(H^2 L)^{\frac{2}{3}}}  \, \left [\frac{c_{8}}{Pm \left( 1 +  \ln Re +  \ln  \frac{L}{\ell}   \right)} \right]^{\frac{2}{3}} \, .
\end{equation}
Neglecting the log term on the right-hand-side and assuming that the geometrical prefactor is of order unity, the maximum Reynolds number for which we can prove two-dimensionalization is roughly $Pm^{-2/3}$. For a liquid metal $Pm \simeq 10^{-6}$, so this is typically $10^4$. Once again, in shallow domains (small $H/\ell$) two-dimensionalization can be proven up to higher $Re$.

\subsection{Kolmogorov forcing}

The tighter upper bound (\ref{2DboundKolmogorov}) on the time-averaged vorticity gradients holds if the forcing function is of single-mode type, that is if $\Delta \textbf{f}$ is proportional to $\textbf{f}$. 
For such forcing we combine (\ref{2DboundKolmogorov}) with equation (\ref{bndLinfty}) to get a bound on the velocity gradients in terms of the rms velocity 
\begin{eqnarray}
\left< \|\bnabla  \textbf{V}\|_\infty \right>  & \leq & c_{9} \frac{U L}{\ell^2} \, \left( 1 + \ln Re + \ln \frac{L}{\ell}  \right) \, ,\end{eqnarray}
which we insert into (\ref{criterion}) to obtain criteria for two-dimensionalization in terms of the rms velocity only:
\begin{eqnarray}
N  & \geq & c_{10} \frac{H^2 L^2}{\ell^4} Re \left( 1 + \ln Re + \ln \frac{L}{\ell}  \right)^2 \, , \label{condition1Kolm}\\
\text{and } Rm & \leq & \frac{\ell^3}{H^2 L} \, \frac{c_{11}}{ \left( 1 + \ln Re + \ln \frac{L}{\ell}  \right) } \, , \label{condition2Kolm}
\end{eqnarray}
where we also assumed $Re \geq 1$ and $Pm \leq 1$ along the way, and $c_{10}$ to $c_{11}$ are numbers which depend on the shape of the forcing only. We can see that the assumption of single-mode forcing allows to prove two-dimensionalization in a larger region of parameter space. Indeed, forgetting about the logarithmic terms, conditions (\ref{condition1Kolm}) and (\ref{condition2Kolm}) indicate that if the interaction parameter is larger than the Reynolds number and the magnetic Reynolds number is smaller than a constant, then the flow remains two-dimensional. Once again, the geometrical prefactors in (\ref{condition1Kolm}) and (\ref{condition2Kolm}) are such that the sufficient conditions for two-dimensionalization are more easily met in shallow domains. Condition (\ref{condition2Kolm}) can be recast in terms of the kinetic Reynolds number only as
\begin{equation}
Re \leq \frac{1}{Pm} \, \frac{\ell^3}{H^2 L} \,  \frac{c_{11}}{ \left( 1 + \ln Re + \ln \frac{L}{\ell}  \right) }  \, . \label{condition2KolmRe}
\end{equation}
For Kolmogorov forcing, we can therefore prove two-dimensionalization up to a Reynolds number of order $Pm^{-1}$. For a liquid metal, this is typically $10^6$. This maximum value of $Re$ for which we can prove two-dimensionalization is larger in shallow domains because of the prefactor $\ell^3/(H^2 L)$.

\section{Discussion\label{section6}}

\begin{table}
\begin{center}
\begin{tabular*}{0.95\textwidth}{@{\extracolsep{\fill}}  c | c | c  }
System & Region of parameter space & Property\\
 & (asymptotic in $Re$) & \\
\hline
Full MHD, time-independent forcing & $\displaystyle N \geq  \text{const} \times \frac{H^2 L^2}{\ell^4} Re^2 [\ln(Re)]^2$ & L2D \\
 & $\displaystyle Rm \leq \text{const} \times \frac{\ell^3}{H^2 L \sqrt{Re} \ln(Re)} $ & \\
 \hline
 Full MHD, single-mode, time-independent forcing & $\displaystyle N \geq  \text{const} \times \frac{H^2 L^2}{\ell^4} Re [\ln(Re)]^2$ & L2D \\
 & $\displaystyle Rm \leq \text{const} \times \frac{\ell^3}{H^2 L \ln(Re)} $ & \\
 \hline
Quasi-static MHD, time-independent forcing & $\displaystyle N \geq  \text{const} \times \frac{H^2 L^4}{\ell^6} Re^5 $ & A2D \\
 \hline
\end{tabular*}
\caption{Regions of parameter space where we proved two-dimensionalization. The threshold values are written in the limit of large Reynolds number. L2D stands for ``linear two-dimensionalization" and A2D for ``absolute two-dimensonalization". \label{table1}}
\end{center}
\end{table}

\begin{figure}
\begin{center}
\includegraphics[width=120 mm]{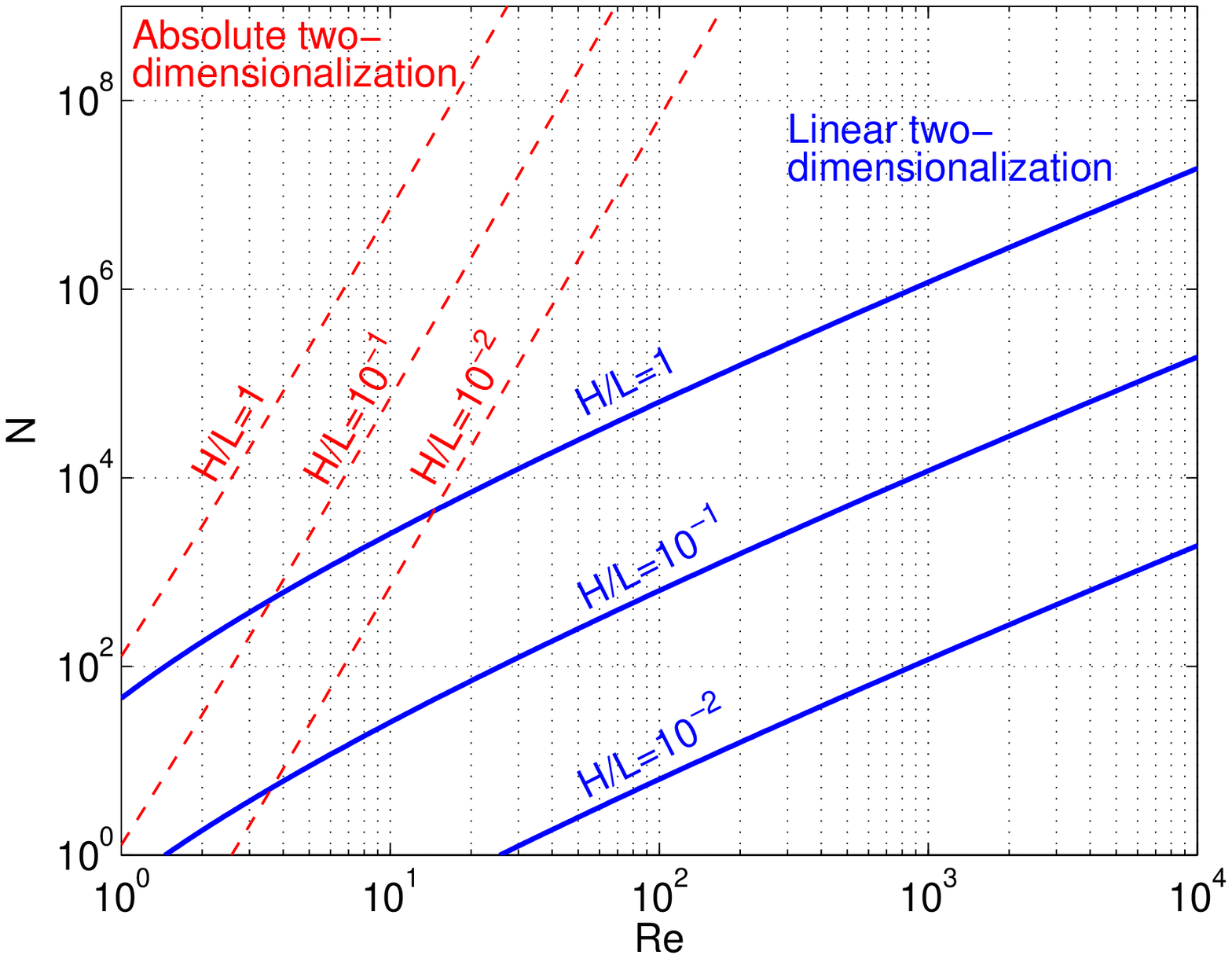}
\includegraphics[width=120 mm]{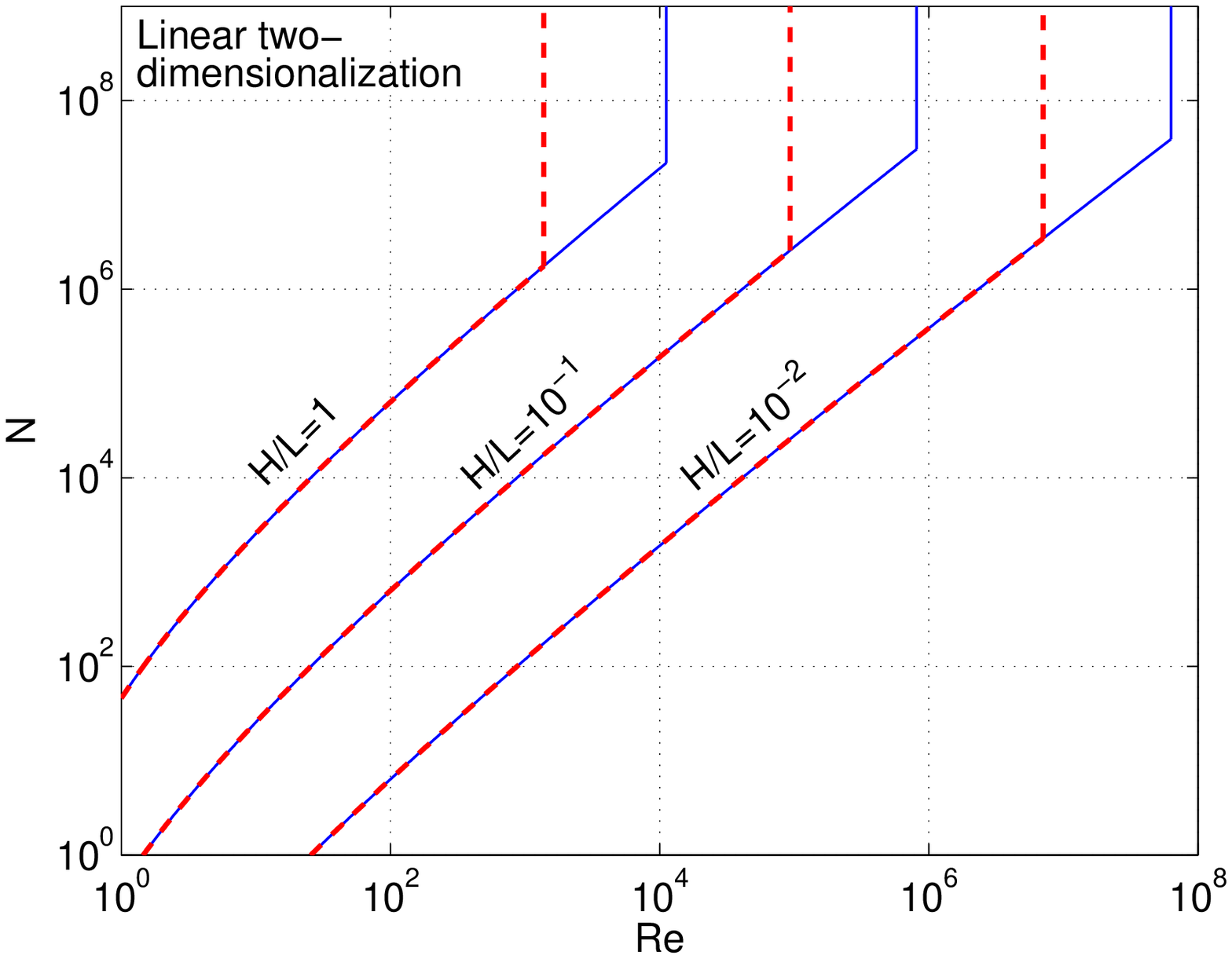}
\end{center}
\caption{A sketch of the bounds (\ref{NcAbsolute}), (\ref{condition1Kolm}) and (\ref{condition2KolmRe}), in parameter space, for single-mode forcing, $\ell=L/2$, and several values of the aspect ratio $H/L$. For simplicity, the dimensionless constants appearing in these expressions have been taken equal to unity. \textbf{Top:} Results obtained in the quasi-static limit. Above the linear two-dimensionalization line (blue, solid line), we have proven that the flow has a stable 2D attractor. However, this attractor is stable for small 3D perturbations only, and fully 3D attractors of the system may exist as well. Above the absolute two-dimensionalization line (red dashed), the flow has only 2D attractors. \textbf{Bottom:} Linear two-dimensionalization for the full-MHD system, for magnetic Prandtl number $Pm=10^{-6}$ (blue solid) and $Pm=10^{-5}$ (red dashed): on the top left part of each curve, we have proven that the system has a purely two-dimensional attractor.\label{scalingQS}}
\end{figure}

In this study we considered a flow driven by a horizontal body-force and subject to a strong vertical magnetic field. We determined regions of parameter space where the two-dimensional solutions of the Navier-Stokes equation are stable with respect to three-dimensional perturbations. Such two-dimensionalization occurs for strong enough interaction parameter $N$ and low enough magnetic Reynolds number $Rm$. We first focused on the $Rm \ll 1$ asymptotic limit, the so-called quasi-static limit, to prove absolute two-dimensionalization: above some $Re$-dependent value of the interaction parameter, the attractors of the flow are all 2D, i.e., the flow becomes 2D in the long-time limit, regardless of the initial condition. We then came back to the full-MHD problem, for which we derived a value $N_c(Re)$ of the interaction parameter and a value $Rm_c(Re)$ of the magnetic Reynolds number such that, if $N\geq N_c$ and $Rm \leq Rm_c$, the 2D flow is stable with respect to infinitesimal vertically-dependent perturbations: we called this property linear two-dimensionalization. Of course, the results proven using the full-MHD equations are also valid in the quasi-static limit $Rm \to 0$: then only $N\geq N_c$ is required for linear two-dimensionalization. Alternatively, one can directly prove linear two-dimensionalization for quasi-static MHD: bound the triple velocity product in (\ref{defH}) by $\| \bnabla {\bf V} \|_\infty \| {\bf v} \|_2^2$, where ${\bf V}$ is now a solution of the 2D Navier-Stokes equation. The dissipative terms are bounded as in (\ref{Fourierineq}), hence $\mathcal{H}\{\textbf{v},t\} \leq ( \| \bnabla {\bf V} \|_\infty -\sqrt{2 \beta \nu}\pi/H)  \| {\bf v} \|_2^2$. Asking for the time-average of the parenthesis to be negative leads to the condition $\frac{\sqrt{\beta \nu}}{H \la \| \bnabla {\bf V} \|_\infty \ra} \geq \frac{1}{\sqrt{2} \pi}$ for linear two-dimensionalization. Conditions expressed in terms of the rms velocity $U$ are obtained using (\ref{bndLinfty}) followed by (\ref{2Dbound}) or (\ref{2DboundKolmogorov}): this gives the same criteria as (\ref{condition1}) and (\ref{condition1Kolm}), albeit with different numerical constants.

The different regions of two-dimensionalization are summarized in Table \ref{table1}, and sketched in figure \ref{scalingQS} in the case of single-mode forcing. These results have implications for the existence of an energy dissipation anomaly in this system. 
Indeed, as can be seen in top panel of figure \ref{scalingQS}, depending on how one takes the double limit $(Re\to \infty, N \to \infty)$, the quasi-static system may or may not exhibit a dissipation anomaly. If we take this limit with $N>Re^5$, then in the long-time limit the flow settles to a purely 2D attractor, with vanishing Ohmic dissipation, and a time-averaged viscous dissipation that follows a laminar scaling, proportional to $Re^{-1}$: there is no dissipation anomaly (on a purely mathematical basis, our proof of absolute two-dimensionalization shows that the 3D part of the flow decays in the $L_2$ sense only, and one could argue that the 3D velocity gradients can remain large. However, physical intuition suggests that the 3D velocity gradients decay when the $L_2$ norm of the 3D velocity decays).

In the region $N>Re (\ln Re)^2$, we know that the system has a 2D attractor that is stable with respect to infinitesimal perturbations. If the initial condition is close enough to being 2D, the system will end up in this purely 2D state, with no energy dissipation anomaly. However, this attractor is not necessarily the only one of the dynamics: if the initial condition is strongly 3D, the flow may end up in a fully-3D attractor, and this 3D state may exhibit a dissipation anomaly.

The limiting values of the interaction parameter derived in the present study are rather conservative: the best criterion for two-dimensionalization we produced is roughly $N \geq Re (\ln Re)^2$, when several examples of 2D flows become stable as soon as $N$ is larger than some $Re$-independent value (see for instance the study of the strained elliptical vortex by \citet{Thess}). It could be that the computed threshold values of $N$ are not sharp, or that the 2D turbulent flows we considered are much more unstable to 3D perturbations than the smooth ideal 2D flows considered by previous authors.

We now wish to comment on the shape of the domain: we see that all the threshold values of $N$  in Table \ref{table1} are proportional to the squared aspect ratio $H^2/L^2$ of the domain, and the threshold $Rm$ are inversely proportional to it: the shallower the domain, the larger the region of parameter space in which we can prove two-dimensionalization, as can be seen in figure \ref{scalingQS}. That is because shallow domains prohibit the long-vertical-wavelength perturbations that are poorly affected by magnetic damping.

Further comments are needed regarding the boundary conditions considered in this study. 
We first stress the fact that the results of this study carry over to 3D periodic domains (albeit with different dimensionless numerical constants), which are more fashionable among numericists. As far as laboratory experiments are concerned, the simple magnetic boundary conditions in (\ref{BC}) can be relevant: they correspond to boundaries with a much higher magnetic permeability than the fluid. As an example, electromagnets often contain iron pieces to channel the magnetic field. If liquid metal is placed in the gap of such an electromagnet, between two iron pieces, the boundaries have very large magnetic permeability provided the ferromagnetic material is not saturated: the magnetic boundary conditions then approach those in (\ref{BC}). Recently, such large-magnetic-permeability boundaries have attracted attention in the dynamo community, because in some geometries they lower the threshold for dynamo action \citep{Gissinger, Gallet2012, Gallet2013, Herault}. However, these are not the most common boundaries: many experiments  and astrophysical situations correspond to non-magnetic and insulating boundaries. These alternate boundary conditions do not affect the results obtained here in the quasi-static approximation, because equation (\ref{NSzeroRm}) still applies. Indeed, both the high-magnetic-permeability and the insulating boundary conditions impose a vanishing normal component of the current density at the boundary, and therefore Neumann boundary conditions for the inverse Laplacian in equation (\ref{NSzeroRm}). As a result, both absolute and linear two-dimensionalization can be proven for insulating boundaries in the quasi-static limit, with the same criteria as for high-magnetic-permeability boundaries. 
 However, the proof of linear two-dimensionalization for the full-MHD equations does not carry over to non-magnetic boundary conditions, although we believe similar results probably hold.

Finally, although the stress-free or 3D-periodic boundary conditions on the velocity field might have some relevance in an astrophysical context, they are a crude approximation to liquid metal experiments, where the fluid flows inside a rigid container. Such no-slip boundaries cannot be taken into account in the present analysis, because they prevent the existence of a purely 2D state. Instead, for strong enough magnetic field the flow settles in a quasi-2D state, with vertically invariant flow in the bulk and sharp Hartmann layers at the horizontal boundaries \citep{Sommeria, Klein, Potherat2014}. These layers can induce additional instabilities \citep{Lingwood, Moresco}, and dissipation through Hartmann friction, which we avoid by considering stress-free or periodic boundary conditions.



Some of this work was performed at the Geophysical Fluid Dynamics Program at Woods Hole Oceanographic Institution, supported by the US National Science Foundation (NSF) Awards OCE-0824636 and OCE-1332750 and the Office of Naval Research.  And some of this work was completed at the NSF Institute for Pure \& Applied Mathematics at UCLA.  This research was also supported in part by NSF Award PHY-1205219 (CRD) and by the junior grant TURBA from Labex PALM ANR-10-LABX-0039 (BG).

\appendix

\section{Extension of the bound on time-averaged viscous energy dissipation to quasi-static MHD\label{appbound}}

\citet{Doering} produced an upper bound on the time-averaged viscous energy-dissipation rate for body-forced 3D turbulence in a periodic domain. We prove here that the same upper bound holds for MHD in the quasi-static limit. Following these authors, we write the forcing function as $\textbf{f}=F\boldsymbol{\phi}(\textbf{x}/\ell) $, where the shape-function $\boldsymbol{\phi}$ is periodic of period $1$ in its two-dimensionless variables $x/\ell$ and $y/\ell$. Its rms value is equal to unity. $\boldsymbol{\phi}$ is the shape of the force.

To compute the kinetic energy budget, we take the dot product of (\ref{NSzeroRm}) with ${\bf u}$ before integrating over the entire domain, which yields
\begin{equation}
\frac{\mathrm{d}}{\mathrm{d}t}\left( \frac{\| {\bf u}\|_2^2}{2}\right) = \int_\mathcal{D} \textbf{f} \cdot \textbf{u} \, \mathrm{d}^3 \textbf{x} - \nu \| \bnabla {\bf u}\|_2^2  - \beta \| \bnabla^{-1} \partial_z {\bf u}\|_2^2 \, .
\end{equation}
After time-averaging, we obtain the following upper bound on the time-averaged viscous energy dissipation rate,
\begin{equation}
\nu \la \| \bnabla {\bf u}\|_2^2 \ra  =  \la \int_\mathcal{D} \textbf{f} \cdot \textbf{u} \mathrm{d}^3 \textbf{x} \ra  - \beta \la \| \bnabla^{-1} \partial_z {\bf u}\|_2^2 \ra \leq   F U L^2 H \, ,
\label{boundepsilontemp}
\end{equation}
where we discarded the second term and used Cauchy-Scwhartz inequality to bound the first one.

To bound the forcing intensity $F$ in terms of the rms velocity $U$, we take the dot product of (\ref{NSzeroRm}) with ${\bf f}$, integrate over the domain and time-average. The contribution from the Joule dissipative term vanishes after an integration by parts to move a $z$-derivative onto the $z$-independent force $\textbf{f}$. Further integrations by parts using the divergence-free constraint lead to
\begin{eqnarray}
\la \|  {\bf f}\|_2^2 \ra = L^2 H F^2 & \leq &   \left| \la   \int_\mathcal{D} \textbf{u} \cdot (\bnabla \textbf{f}) \cdot \textbf{u} \, \mathrm{d}^3 \textbf{x}  \ra  \right|  + \nu \left| \la   \int_\mathcal{D} \textbf{u} \cdot \Delta \textbf{f} \, \mathrm{d}^3 \textbf{x}  \ra  \right| \\
\nonumber & \leq & \tilde{c}_1 \frac{F}{\ell} L^2 H U^2 + \tilde{c}_2 \frac{\nu F L^2 H U}{\ell^2} \, ,
\end{eqnarray}
where $\tilde{c}_1$ and $\tilde{c}_2$ are dimensionless positive numbers that depend on the shape ${\boldsymbol \phi}$ of the forcing only. Multiplying by $U/F$ yields
\begin{eqnarray}
L^2 H F U & \leq & \tilde{c}_1 \frac{U^3}{\ell} L^2 H + \tilde{c}_2 \frac{\nu L^2 H U^2}{\ell^2} \, ,
\end{eqnarray}
which we use to bound the right-hand-side of the inequality (\ref{boundepsilontemp}). This gives a bound that is similar to \citet{Doering},
\begin{equation}
\nu \la \| \bnabla {\bf u}\|_2^2 \ra  \leq  \tilde{c}_1 \frac{U^3}{\ell} L^2 H + \tilde{c}_2 \frac{\nu L^2 H U^2}{\ell^2} =  L^2 H \frac{U^3}{\ell} \left( \tilde{c}_1+ \frac{\tilde{c}_2}{Re} \right)\, .
\end{equation}

\section{Bounding the triple velocity product using Ladyzhenskaya's inequality \label{appendix1}}

\subsection{Ladyzhenskaya's inequality in a 2D periodic box}


For this subsection, we consider functions of $x$ and $y$ only, and we denote as  $\| \dots \|_{2;\perp}$ the standard $L_2$ norm in 2D.  Ladyzhenskaya's inequality provides a bound on the $L_4$ norm of some field in terms of the $L_2$ norms of the field and its gradient. For a zero-mean scalar field $\tilde{f}(x,y)$ in the periodic two-torus $\mathcal{T}_2=[0,L]^2$, it reads
\begin{equation}
\iint_{\mathcal{T}_2} \tilde{f}^4 \mathrm{d}x \mathrm{d}y \leq \mathcal{C} \| \bnabla \tilde{f} \|_{2;\perp}^2 \| \tilde{f} \|_{2;\perp}^2 \, ,
\end{equation}
where $\mathcal{C}$ is a dimensionless constant of order unity. 

Let us consider a scalar field $f(x,y)$ with a nonzero mean $F$ over $\mathcal{T}_2$ and a fluctuating part $\tilde{f}(x,y)$, such that $f(x,y)=F+\tilde{f}(x,y)$. We may bound the $L_4$ norm of $f(x,y)$ according to
\begin{eqnarray}
\iint_{\mathcal{T}_2} f^4 \mathrm{d}x \mathrm{d}y & \leq & F^4 L^4 + 8 F^2 \| \tilde{f} \|_{2;\perp}^2 + 3 \iint_{\mathcal{T}_2} \tilde{f}^4 \mathrm{d}x \mathrm{d}y \\
\nonumber & \leq & \frac{3}{L^2} \| f \|_{2;\perp}^4 + 3 \iint_{\mathcal{T}_2} \tilde{f}^4 \mathrm{d}x \mathrm{d}y \\
\nonumber & \leq & \frac{3}{L^2} \| f \|_{2;\perp}^4 + 3 \mathcal{C}  \| \bnabla \tilde{f} \|_{2;\perp}^2 \| \tilde{f} \|_{2;\perp}^2 \\
\nonumber & \leq & 3 \left( \frac{1}{L^2} + \frac{\mathcal{C}}{2a L^2} \right) \| f \|_{2;\perp}^4 + \frac{3 \mathcal{C} a L^2}{2}  \| \bnabla f \|_{2;\perp}^4 \, ,
\end{eqnarray}
where we used Ladyzhenskaya's inequality for the third step and Young's inequality for the fourth step, with $a$ some arbitrary positive number. Taking the square root of the last inequality and using a convexity inequality, we obtain
\begin{equation}
\sqrt{ \iint_{\mathcal{T}_2} f^4 \mathrm{d}x \mathrm{d}y } \leq \frac{1}{L}   \sqrt{3 + \frac{3\mathcal{C}}{2a} } \| f \|_{2;\perp}^2 + L \sqrt{ \frac{3 \mathcal{C} a}{2} }\| \bnabla f \|_{2;\perp}^2 \, , \label{Ladynonzeromean}
\end{equation}
where the positive number $a$ can be chosen to optimize the bound.

\subsection{Bound on the triple velocity product}


Let us write the three-dimensional part of the velocity field ${\bf v}=(u,v,w)$, and introduce Fourier series in $z$ for the horizontal components $u$ and $v$:
\begin{eqnarray}
u(x,y,z,t) & = & \sum_{p=1}^{p=+\infty} u_p(x,y,t) \cos \left( p \pi \frac{z}{H} \right) \, , \\
v(x,y,z,t) & = & \sum_{p=1}^{p=+\infty} v_p(x,y,t) \cos \left( p \pi \frac{z}{H} \right) \, . \\
\end{eqnarray}
The $L_2$ norm of one such horizontal component is $\| u \|_2^2 = \frac{H}{2} \sum_{p=1}^{p=+\infty} \| u_p \|_{2;\perp}^2$.

Using these Fourier series in the vertical together with the Cauchy-Schwarz inequality, recalling that ${\bf V}$ is independent of $z$, we bound the triple velocity product according to
\begin{eqnarray}
 \left| \int_\mathcal{D} \textbf{v} \cdot (\bnabla \textbf{V}) \cdot \textbf{v} \, \mathrm{d}^3 \textbf{x} \right| \leq \frac{H}{2}  \| \bnabla {\bf V} \|_{2;\perp} \sum_{p=1}^{p=+\infty} \left[ \sqrt{ \iint_{\mathcal{T}_2} u_p^4 \mathrm{d}x \mathrm{d}y } +  \sqrt{ \iint_{\mathcal{T}_2} v_p^4 \mathrm{d}x \mathrm{d}y }   \right] \, ,
\end{eqnarray}
and using inequality (\ref{Ladynonzeromean}) for $u_p$ and $v_p$, we obtain
\begin{eqnarray}
 \left| \int_\mathcal{D} \textbf{v} \cdot (\bnabla \textbf{V}) \cdot \textbf{v} \, \mathrm{d}^3 \textbf{x} \right| & \leq & \| \bnabla {\bf V} \|_{2;\perp} \left[  \frac{1}{L} \sqrt{3  + \frac{3 \mathcal{C}}{2a} } (\| u \|_{2;\perp}^2 + \| v \|_{2;\perp}^2 ) \right. \\
 \nonumber & & \left. +  L \sqrt{ \frac{3 \mathcal{C} a}{2} } ( \| \bnabla u \|_{2;\perp}^2 + \| \bnabla v \|_{2;\perp}^2 )  \right]  \\
 \nonumber & \leq & \frac{\| \bnabla \textbf{V} \|_2}{\sqrt{H}} \left[  \frac{1}{L} \sqrt{3 + \frac{3 \mathcal{C}}{2a} } \| {\bf v} \|_2^2  +L \sqrt{ \frac{3 \mathcal{C} a}{2} } \| \bnabla {\bf v} \|_2^2 \right] \, .
\end{eqnarray}
In this expression the positive number $a$ can be chosen arbitrarily to optimize the bound, and we set $a=\frac{\nu^2 H}{6 {\cal C}L^2 \| \bnabla \textbf{V} \|_2^2}$. Using the concavity of the square-root function we finally obtain
\begin{eqnarray}
 \left| \int_\mathcal{D} \textbf{v} \cdot (\bnabla \textbf{V}) \cdot \textbf{v} \, \mathrm{d}^3 \textbf{x} \right| & \leq & \frac{\| \bnabla \textbf{V} \|_2}{\sqrt{H}} \left[  \frac{1}{L} \left( \sqrt{3} + \sqrt{\frac{3 \mathcal{C}}{2a}}   \right) \| {\bf v} \|_2^2  +L \sqrt{ \frac{3 \mathcal{C} a}{2} } \| \bnabla {\bf v} \|_2^2 \right] \, ,\\
 \nonumber & \leq & c \left( \frac{ \| \bnabla \textbf{V} \|_2}{L \sqrt{H}} + \frac{\| \bnabla \textbf{V} \|_2^2}{\nu H}  \right) \| {\bf v} \|_2^2 + \frac{\nu}{2}  \| \bnabla {\bf v} \|_2^2 \, .
\end{eqnarray}

\section{Bound on the time-average of the $L^\infty$ norm of the velocity gradients\label{Appendix2}}

Throughout this appendix we consider 2D functions of $x$ and $y$ only, and for brevity we denote $\| \dots \|_2$ as the $L_2$ norm in 2D. We wish to bound the time average of the $L^\infty$ norm of the velocity gradients of a solution of the body-forced 2D Navier-Stokes equation. This bound should depend on the rms velocity only.

Decompose $\textbf{V}$ into Fourier components,
\begin{equation}
\textbf{V} = \sum\limits_{\textbf{k}} {\textbf{V}}_\textbf{k}(t) e^{i \textbf{k}\cdot \textbf{x}} + c.c. \, ,
\label{Fourier3D}
\end{equation}
where the wavevector $\textbf{k}$ takes the values 
\begin{equation}
\textbf{k}=(k_x,k_y)=\left(n_x \frac{2\pi}{L}, n_y \frac{2\pi}{L} \right), \mbox{with } {(n_x,n_y) \in \mathbb{N}\times\mathbb{N}} 	\setminus (0,0) \, .
\end{equation}
Consider an arbitrary cut-off wavenumber $Q>0$ and write
\begin{eqnarray}
 \| \bnabla {\bf V} \|_\infty & \leq & \sum\limits_{{\bf k}} k |{\bf V}_{\bf k}| = \sum\limits_{k < Q} k |{\bf V}_{\bf k}| + \sum\limits_{k \geq Q} k |{\bf V}_{\bf k}| \\
 \nonumber & \leq & \sqrt{\sum\limits_{k < Q} k^4 |{\bf V}_{\bf k}|^2} \sqrt{\sum\limits_{k < Q} \frac{1}{k^2}} + \sqrt{\sum\limits_{k \geq Q} k^6 |{\bf V}_{\bf k}|^2} \sqrt{\sum\limits_{k \geq Q} \frac{1}{k^4}}
\end{eqnarray}
Now the sums that are independent of ${\bf V}$ can be bounded by integrals, with the following result
\begin{eqnarray}
\sum\limits_{k < Q} \frac{1}{k^2} & \lesssim & L^2 \ln Q \, , \\
\nonumber  \sum\limits_{k \geq Q} \frac{1}{k^4} & \lesssim & \frac{L^2}{Q^2} \, ,
\end{eqnarray}
where the notation $a \lesssim b$ means that there exists $c >0$ such that $a \leq c b$, where $c$ is a dimensionless positive constant that is independent of the parameters of the problem.
Hence
\begin{eqnarray}
 \| \bnabla {\bf V} \|_\infty & \lesssim & \frac{L}{Q} \| \Delta \omega \|_2 + \sqrt{\ln Q} \| \bnabla \omega \|_2 \, .
\end{eqnarray}
Upon picking the value $Q=L \| \Delta \omega \|_2/\sqrt{D}$, where $D=\| \bnabla \omega \|_2^2+ \left< \| \bnabla \omega \|_2^2 \right>$ is a strictly positive time-dependent quantity -- this value of $D$ allows for the following simple upper bound on $D^{-1}$, $D^{-1}\leq \left< \| \bnabla \omega \|_2^2 \right>^{-1}$, which is needed to obtain inequality (\ref{boundforcingpalin}) -- and time-averaging, we obtain
\begin{eqnarray}
\la \| \bnabla {\bf V} \|_\infty \ra & \lesssim & \la \sqrt{D} \ra + \la \frac{1}{2} \| \bnabla \omega \|_2 \sqrt{ \ln{\frac{L^2 \| \Delta \omega \|_2^2}{D}}   } \ra \, ,
\end{eqnarray}
and using Young's and Jensen's inequalities,
\begin{eqnarray}
\la \| \bnabla {\bf V} \|_\infty \ra & \lesssim &  \sqrt{\la \| \bnabla \omega \|_2^2 \ra}  \left( 1+ \ln \la \frac{L^2 \| \Delta \omega \|_2^2}{D}  \ra  \right)  \, . \label{inftytemp}
\end{eqnarray}

We now bound the argument of the logarithm. Our goal is only to prove that it has at most an algebraic dependence on Reynolds number, so that the logarithmic term admits a bound of the form $\text{const} \times \ln(Re)$. The 2D Navier-Stokes equation written for the vorticity is
\begin{equation}
\partial_t \omega + \textbf{V} \cdot \bnabla \omega = \nu \Delta \omega + (\bnabla \times \textbf{f})\cdot \textbf{e}_z
\end{equation}
Upon multiplying by $-\Delta \omega$ and integrating over space, we get the palinstrophy evolution equation
\begin{equation}
\frac{\mathrm{d}}{\mathrm{d}t} \left( \frac{\| \bnabla \omega \|_2^2}{2} \right) = \int_{[0,L]^2} (\textbf{V} \cdot \bnabla \omega) \Delta \omega \, \mathrm{d} \textbf{x} - \nu \| \Delta \omega  \|_2^2 - \int_{[0,L]^2} (\bnabla \times \omega \textbf{e}_z) \cdot \Delta \textbf{f} \, \mathrm{d} \textbf{x}
\end{equation}
Division by $D$ yields
\begin{equation}
\frac{1}{2} \frac{\mathrm{d}}{\mathrm{d}t} \left( \ln D \right) = \frac{1}{D }\int_{[0,L]^2} (\textbf{V} \cdot \bnabla \omega) \Delta \omega \, \mathrm{d} \textbf{x} - \nu \frac{ \| \Delta \omega  \|_2^2}{D} - \frac{1}{D} \int_{[0,L]^2} (\bnabla \times \omega \textbf{e}_z) \cdot \Delta \textbf{f} \, \mathrm{d} \textbf{x} \, ,
\end{equation}
and after time-averaging
\begin{equation}
0 = \left< \frac{1}{D }\int_{[0,L]^2} (\textbf{V} \cdot \bnabla \omega) \Delta \omega \, \mathrm{d} \textbf{x} \right> - \nu \left< \frac{ \| \Delta \omega  \|_2^2}{D} \right> - \left< \frac{1}{D} \int_{[0,L]^2} (\bnabla \times \omega \textbf{e}_z) \cdot \Delta \textbf{f} \, \mathrm{d} \textbf{x} \right>  \, , \label{palinbudget}
\end{equation}
To bound the first term, let us first use the H\"older-Cauchy-Schwarz-Young recipe
\begin{equation}
\left| \int_{[0,L]^2} (\textbf{V} \cdot \bnabla \omega) \Delta \omega \, \mathrm{d} \textbf{x} \right | \leq \| \textbf{V} \|_\infty  \| \bnabla \omega \|_2  \| \Delta \omega \|_2 \leq \frac{\nu}{2} \| \Delta \omega \|_2^2 +\frac{1}{2\nu} \| \textbf{V} \|_\infty^2 \| \bnabla \omega \|_2^2 \, ,
\end{equation}
and divide by $D$ to get
\begin{equation}
\frac{1}{D} \left| \int_{[0,L]^2} (\textbf{V} \cdot \bnabla \omega) \Delta \omega \, \mathrm{d} \textbf{x} \right | \leq \nu \frac{\| \Delta \omega \|_2^2}{2D}  +\frac{1}{2\nu} \| \textbf{V} \|_\infty^2 \, , \label{terme1}
\end{equation}
Since we only want to prove algebraic dependence on Reynolds number, the following simple estimate is sufficient at this stage:
\begin{equation}
\| \textbf{V} \|_\infty^2 \lesssim  \| \textbf{V} \|_2 \| \bnabla \omega \|_2 \lesssim \frac{1}{2 L U} \sqrt{ \left< \| \bnabla \omega \|_2^2 \right>}  \| \textbf{V} \|_2^2 + \frac{ L U}{2  \sqrt{\left< \| \bnabla \omega \|_2^2 \right>}}\| \bnabla \omega \|_2^2 \, .
\end{equation}
The first inequality is a standard Agmon's inequality for vector fields in a 2D periodic domain. After time averaging, we obtain $\left< \| \textbf{V} \|_\infty^2 \right> \lesssim U L \sqrt{\left< \| \bnabla \omega \|_2^2 \right>}$. The time-average of (\ref{terme1}) gives
\begin{equation}
\left< \frac{1}{D} \left| \int_{[0,L]^2} (\textbf{V} \cdot \bnabla \omega) \Delta \omega \, \mathrm{d} \textbf{x} \right| \right> \lesssim \frac{\nu}{2} \left< \frac{\| \Delta \omega \|_2^2}{D} \right>  +\frac{ U L}{2\nu} \sqrt{\left< \| \bnabla \omega \|_2^2 \right>} \, . \label{bound3rd}
\end{equation}
Using Cauchy-Schwarz inequality, together with the convexity inequality $\left< \| \bnabla \omega \|_2 \right> \leq \sqrt{ \left< \| \bnabla \omega \|_2^2 \right>  }$ and Poincar\'e's inequality, the term involving the forcing is bounded by
\begin{eqnarray}
\label{boundforcingpalin} \left< \left|   \frac{1}{D} \int_{[0,L]^2} (\bnabla \times \omega \textbf{e}_z) \cdot \Delta \textbf{f} \, \mathrm{d} \textbf{x}   \right| \right> & \leq & \left< \frac{\| \bnabla \omega \|_2 \| \Delta \textbf{f} \|_2}{D} \right > \\
\nonumber & \leq &   \left< \frac{\| \bnabla \omega \|_2 \| \Delta \textbf{f} \|_2}{\left< \| \bnabla \omega \|_2^2 \right>} \right>  \leq \frac{\| \Delta \textbf{f} \|_2}{\sqrt{\left< \| \bnabla \omega \|_2^2 \right>}} \leq \frac{L \| \Delta \textbf{f} \|_2}{\pi^2 U} \, .
\end{eqnarray}
Recall that we write the forcing function as $\textbf{f}=F\boldsymbol{\phi}(\textbf{x}/\ell) $, where $\boldsymbol{\phi}$ is periodic of period $1$ in its two dimensionless variables and has an rms value $1$. Then following Alexakis and Doering, we can bound $F$ in terms of the rms velocity $U$: their equation (18) gives
\begin{equation}
F \leq \frac{U^2}{\ell} \left( \tilde{c}_1 +\frac{\tilde{c}_2}{Re}\right) \, , \text{ hence } \| \Delta \textbf{f} \|_2 \leq \frac{U^2 L}{\ell^3}  \left( \tilde{c}_3 +\frac{\tilde{c}_4}{Re}\right) \, , \label{boundforcing}
\end{equation}
where the $(\tilde{c}_i)_{i\in \mathbb{N}}$ appearing throughout this appendix are dimensionless positive numbers that depend on the shape ${\boldsymbol \phi}$ of the forcing only.

Upon inserting (\ref{bound3rd}), (\ref{boundforcingpalin}) and (\ref{boundforcing}) into (\ref{palinbudget}), we obtain
\begin{eqnarray}
\left< \frac{ L^2 \| \Delta \omega  \|_2^2}{D} \right>  & \lesssim &   Re \frac{L^3 \sqrt{\left< \| \bnabla \omega \|_2^2 \right>}}{\ell \nu} + \frac{ L^4 U}{\ell^3 \nu} \left(  \tilde{c}_3 +\frac{\tilde{c}_4}{Re} \right) \, , \\
& \lesssim &  Re \frac{L^3 \sqrt{\left< \| \bnabla \omega \|_2^2 \right>}}{\ell \nu} + \frac{ L^4}{ \ell^4}  \left(  \tilde{c}_3 Re + \tilde{c}_4\right) \, ,
\end{eqnarray}
which is inserted into (\ref{inftytemp}) to get an upper bound on the infinite norm of the velocity gradients in terms of the $L_2$ norm of the vorticity gradients
\begin{eqnarray}
\nonumber \left< \|\bnabla  \textbf{V}\|_\infty \right> & \lesssim &  \sqrt{ \left< \| \bnabla \omega \|_2^2 \right>} \left[ 1+ \ln \left(1+ Re \frac{L^3 \sqrt{\left< \| \bnabla \omega \|_2^2 \right>}}{\ell \nu} + \frac{L^4}{ \ell^4}  \left(  \tilde{c}_3 Re + \tilde{c}_4\right)  \right)  \right] \\
 \label{bndtemp}
\end{eqnarray}
Since bounds on $\left< \| \bnabla \omega \|_2^2 \right>$ can be computed in terms of the rms velocity for body-forced two-dimensional turbulence \citep{Alexakis}, the right-hand-side of (\ref{bndtemp}) can be expressed in terms of the rms velocity only. These bounds show that the argument of the logarithm has at most an algebraic dependence in $Re$, hence, for $Re\geq 1$,

\begin{eqnarray}
 \left< \|\bnabla  \textbf{V}\|_\infty \right> & \leq &  \tilde{c}_5 \sqrt{ \left< \| \bnabla \omega \|_2^2 \right>} \left( 1+ \ln Re + \ln \frac{L}{\ell}  \right)  \, ,  \label{bndinftyapp}
\end{eqnarray}
where $\tilde{c}_5$ depends on the shape of the forcing only. This constant is denoted as $c_5$ in the main body of this article, where the 2D $L_2$ norm $\| \bnabla \omega \|_2$ of this appendix is replaced by its 3D counterpart.

\end{document}